\documentclass[11pt]{article}

\begin{document}

\date{}

\title{Quantum Cosmology and Tachyons}

\author{D. D. Dimitrijevi\'c, G. S. Djordjevi\'c and Lj. Ne\v si\'c\\
Department of Physics, University of Ni\v s\\
 P.O. Box 224, 18000 Ni\v s, Serbia\\
 e-mail: gorandj@junis.ni.ac.yu}

\maketitle

\begin{abstract}

We discuss the relevance of the classical and quantum rolling
tachy\-ons inflation in the frame of the standard, $p$-adic and
adelic minisuperspace quantum cosmology. The field theory of tachyon
matter proposed by Sen in a zero-dimensional version suggested by
Kar leads to a model of a particle moving in a constant external
field with quadratic damping. We calculate the exact quantum
propagator of the model, as well as, the vacuum states and
conditions necessary to construct an adelic generalization.

\end{abstract}

\section{Introduction}
The main task of quantum cosmology \cite{wiltshire} is to describe
the evolution of the universe in a very early stage. Usually one
takes the universe is described by a complex-valued wave function.
Since quantum cosmology is related to the Planck scale phenomena it
is logical to consider various geometries (in particular
nonarchimedean \cite{we-vol} and noncommutative \cite{GOR} ones) and
parametrizations of the space-time coordinates: real, $p$-adic, or
even adelic \cite{vvz}. In this article, we will generally maintain
space-time coordinates and matter fields to be real and $p$-adic.

Supernova Ia observations show that the expansion of the Universe
is accelerating \cite{perlmuter}, contrary to
Friedmann-Robertson-Walker (FRW) cosmological models, with
non-relativistic matter and radiation. Also, cosmic microwave
background (CMB) radiation data are suggesting that the expansion
of our Universe seems to be in an accelerated state which is
referred to as the ``dark energy`` effect. The cosmological
constant as the vacuum energy can be responsible for this
evolution by providing a negative pressure. A need for
understanding these new and rather surprising facts, including
(cold) ``dark matter``, has motivated numerous authors to
reconsider different inflation scenarios. Despite some evident
problems \cite{sami} such as a non-sufficiently long period of
inflation, tachyon-driven scenarios \cite{gibbons, ghoshal, GO}
remain highly interesting for study.

There have been a number of attempts to understand this description
of the early Universe via (classical) nonlocal cosmological models,
first of all via $p$-adic inflation models \cite{barnaby,
jokuovskaya}, which are represented by nonlocal $p$-adic string
theory coupled to gravity. For these models, some rolling
inflationary solutions were constructed and compared with CMB
observations. Another example is the investigation of the $p$-adic
inflation near a maximum of the nonlocal potential when non-local
derivative operators are included in the inflaton Lagrangian. It was
found that higher-order derivative operators can support a
(sufficiently) prolonged phase of slow-roll inflation \cite{lidsey}.

The $p$-Adic approach in HEP is mostly motivated by the following
reasons: (i) the field of rational numbers $Q$, which contains all
observational and experimental numerical data, is a dense subfield
not only in the field of real numbers $R$ but also in the fields
of $p$-adic numbers $Q_p$ ($p$ is a prime number); (ii) there is a
plausible analysis within and over $Q_p$ as well as that one
related to $R$; (iii) general mathematical methods and fundamental
physical laws should be invariant under an interchange of the
number fields \cite{volovich} $R$ and $Q_p$; (iv) there is a
quantum gravity uncertainty while measuring distances around the
Planck length, which restricts the priority of Archimedean
geometry based on real numbers and gives rise to the employment of
non-Archimedean geometry related to $p$-adic numbers; (v) it seems
to be quite reasonable to extend compact Archimedean geometries by
the non-Archimedean ones when integrating over geometries in the
path integral method; and (vi) adelic quantum mechanics applied to
quantum cosmology provides realization of all the above
statements.

In the unified form, adelic quantum mechanics contains ordinary and
all $p$-adic quantum mechanics. As there is not an appropriate
$p$-adic Schr\"odinger equation, there is also no $p$-adic
generalization of the Wheeler-De Witt equation. Instead of the
differential approach, Feynman's path integral method is exploited
\cite{goranbranko}. $p$-Adic gravity and the wave function of the
universe were considered \cite{arefdragvol} as an idea of the
fluctuating number fields at the Planck scale. Like in adelic
quantum mechanics, the adelic eigenfunction of the universe is a
product of the corresponding eigenfunctions of real and all $p$-adic
cases. It was shown that in the framework of this procedure one
obtains an adelic wave function for the de Sitter minisuperspace
model. However, the adelic generalization with the Hartle-Hawking
$p$-adic prescription does not work well when minisuperspace has
more than one dimension, in particular, when matter fields are taken
into consideration. The solution of this problem was found by
treating minisuperspace cosmological models as models of adelic
quantum mechanics. It is a strong motivation to study a class of
exactly solvable quantum mechanical models and apply them in the
frame of quantum cosmology. For the review and detailed discussion
see \cite{we-vol, we6}. The nonarchimedean and noncommutative
cosmological quantum models with extra dimensions and an
accelerating phase have been considered \cite{we2}, as well as the
relevant models and techniques in pure quantum mechanical context
\cite{we1, we3, we4, we5}.

Following S. Kar's \cite{kar} idea on the possibility of the
examination of zero dimensional theory of the field theory of (real)
tachyon matter, and motivated by successes and shortcomings of
classical $p$-adic inflation, we consider real and $p$-adic aspects
of a relevant model with quadratic damping. We calculated the
corresponding propagator and considered vacuum states for $p$-adic
and adelic tachyons.

\section{Quantum Cosmology}

According to the standard cosmological model, in the very
beginning the universe was very small, dense, hot and started to
expand. This initial period of evolution should be described by
quantum cosmology. In the path integral approach to quantum
cosmology over the field of real numbers $R$, the starting point
is the idea that the amplitude to go from one state with intrinsic
metric $h_{ij}^\prime$, and matter configuration $\phi^\prime$ on
an initial hypersurface $\Sigma^\prime$, to another state with
metric $h_{ij}^{\prime\prime}$, and matter configuration
$\phi^{\prime\prime}$ on a final hypersurface
$\Sigma^{\prime\prime}$, is given by a functional integral of the
form
\begin{equation}
\langle h_{ij}'',\phi'',\Sigma''| h_{ij}',\phi',\Sigma'\rangle =
\int {\cal D}{g_{\mu\nu}} {\cal D}\Phi e^{-S[g_{\mu\nu},\Phi]},
\label {4.1}
\end{equation}
over all four-geometries $g_{\mu\nu}$, and matter configurations
$\Phi$, which interpolate between the initial and final
configurations. In this expression $S[g_{\mu\nu},\Phi]$ is an
Einstein-Hilbert action for the gravitational and matter fields
(which can be massless, minimally or conformally coupled with
gravity). This expression stays valid in the $p$-adic case too,
because of its form invariance under change of real to the
$p$-adic number fields.

Among many cosmological models, there is one very important type
of models, the so-called de Sitter models. de Sitter models are
models with the cosmological constant $\Lambda$ and without matter
fields. Models of this type are exactly soluble models and because
of that, they play a role similar to a linear harmonic oscillator
in ordinary quantum mechanics. The general form of the metric for
these models is \cite{we-myers}
\begin{equation}
\label{5.3} ds^2=\sigma^2 [-N^2dt^2+a^2(t)d\Omega^2_{D-1}],
\end{equation}
where $d\Omega^2_{D-1}$ denotes the metric on the unit
$(D-1)$-sphere, $\ \sigma^{D-2}={8\pi G}/$ ${V^{D-1}(D-1)(D-2)}$, $\
V^{D-1}$ is the volume of the unit $(D-1)$-sphere. In the $D=3$
case, this model is related to the multiple sphere configuration and
wormhole solutions. $\upsilon$-Adic ($\upsilon =\infty$ for the
real, and $\upsilon = p$ in the $p$-adic cases) classical action for
this model is
\begin{equation}
\label{5.4} \bar S_\upsilon(a'',N; a',0)= \frac{1}{2\sqrt\lambda}
\left[ N\sqrt\lambda + \lambda \left(
\frac{2a''a'}{\sinh(N\sqrt\lambda)} -
\frac{a'^2+a''^2}{\tanh(N\sqrt\lambda)}\right)\right].
\end{equation}
Let us note that $a$ denotes a scale factor and $\lambda$ denotes
here the appropriately rescaled cosmological constant $\Lambda$,
i.e. $\lambda=\sigma^2\Lambda$. This model was investigated in all
aspects ($p$-adic, real and adelic) in Ref. \cite{we-vol}.
Especially, for this model, the adelic wave function (which unifies
the wave function over the field of real numbers and wave functions
over the field of $p$-adic numbers), is in the form
\begin{equation}
\label{5.5} \Psi(a)=\Psi_\infty(a)\prod_{p}\Psi_p(a_p),
\end{equation}
where $\Psi_\infty(a)$ is a standard wave function and
$\Psi_p(a_p)$ are $p$-adic wave functions. It is very important
that only for finite numbers of $p$, $p$-adic wave functions can
be different from $\Omega$ function which is defined by the
$\Omega (|x|_p) = 1,\ \textrm{for}\ |x|_p \leq 1$ and $ \Omega
(|x|_p) = 0, \ \textrm{for}\ |x|_p
> 1$.

At this place we indicate a considerable similarity of the action
(\ref{5.4}) for the de Sitter model in 2+1 dimensions with the
action (\ref{lj3}) for the tachyon field in the zero dimensional
model, i.e. ``quadratically damped particle under gravity``.

\section{$p$-Adic inflation}

Cosmological inflation has become an integral part of the standard
model of the universe. It provides important clues for structure
formation in the universe and is capable of removing the
shortcomings of the standard cosmology. Gibbons \cite{gibbons} has
emphasized the cosmological implication of tachyonic condensate
rolling towards its ground state. The tachyonic matter might provide
an explanation for inflation at the early epochs and could
contribute to a new form of dark matter at later times.

A recent paper on $p$-adic inflation \cite{barnaby} gives rise to
the hopes that nonlocal inflation can succeed where the real string
theory fails. Starting from the action of the $p$-adic string, with
$m_s$ the string mass scale and $g_s$ the open string coupling
constant,

\begin{equation}
\label{stringaction} S=\frac{m_s^4}{g_p^4}\int d^4 x\left(
-\frac{1}{2}\phi
p^{-\frac{-\partial_t^2+\bigtriangledown^2}{2m_s^2}}\phi+\frac{1}{p+1}\phi
^{p+1} \right), \, \, \,
\frac{1}{g_p^2}=\frac{1}{g_s^2}\frac{p^2}{p-1},
\end{equation}
for the open string tachyon scalar field $\phi (x)$, it has been
shown that a $p$-adic tachyon drives a sufficiently long period of
inflation while rolling away from the maximum of its potential.
Even though this result is constrained by $p\gg 1$ and obtained by
an approximation, it is a good motivation to consider $p$-adic
inflation for different tachyonic potentials. In particular, it
would be interesting to study $p$-adic inflation in quantum regime
and in adelic framework to overcome the constraint $p\gg 1$, with
an unclear physical meaning. For some details, and further
development see \cite{jokuovskaya} and the references therein.

\section{Classical and quantum tachyons}

A. Sen proposed a field theory of tachyon matter a few years ago
\cite{sen} (see also \cite{GO}). The action is given as:
\begin{equation}
  \label{eq:bpu1}
  S=-\int d^{D+1} x V(T) \sqrt{1+\eta^{ij}\partial_i T\partial_j T}
\end{equation}
where $\eta_{00}=-1$ and $\eta_{\alpha\beta}=\delta_{\alpha\beta}$,
$\alpha, \beta=1,2,3,...,D$, $T(x)$ is the scalar tachyon field and
$V(T)$ is the tachyon potential which unusually appears in the
action as a multiplicative factor and has (from string field theory
arguments) exponential dependence with respect to the tachyon field
$V(T)\sim e^{-\alpha T/2}$. In this paper we will focus our
attention on this type of the potential. It is very useful to
understand and to investigate lower dimensional analogs of this
tachyon field theory. The corresponding zero dimensional analogue of
a tachyon field can be obtained by the correspondence:
$x^i\rightarrow t, T\rightarrow x, V(T)\rightarrow V(x)$. The action
reads
\begin{equation}
  \label{eq:bpu3}
  S=-\int dtV(x) \sqrt{1-\dot x^2}.
\end{equation}
In what follows, all variables and parameters can be treated as real
or $p$-adic without a formal change in the obtained forms. It is not
difficult to see that action (\ref{eq:bpu3}), with some appropriate
replacement leads to the equation of motion for a particle with mass
$m$, under a constant external force, in the presence of quadratic
damping:
\begin{equation}
  \label{eq:bpu5}
  m\ddot y + \beta \dot y^2 = mg.
\end{equation}
This equation of motion can be obtained from two Lagrangians
\cite{we-kar}:
\begin{equation}
  \label{eq:eq3}
  L(y,\dot y)=\left (\frac{1}{2}m\dot y^2 +
\frac{m^2g}{2\beta}\right )e^{2{\frac{\beta} {m}}
  y},
  \end{equation}
\begin{equation}
  \label{eq:eq4}
  L(y, \dot y)=-e^{-\frac{\beta}{m} y}\sqrt{1-\frac{\beta}{mg}\dot
  y^2}.
\end{equation}
Despite the fact that different Lagrangians can give rise to
nonequivalent quantization, we will choose the form (\ref{eq:eq3})
that can be handled easily. The first one is better because of the
presence of the square root in the second one. The general solution
of the equation of motion is
\begin{equation} \label{gensol}
y(t)=C_2+\frac{m}{\beta}\ln[\cosh(\sqrt{\frac{g\beta}{m}}\,t+C_1)].
\end{equation}
For the initial and final conditions $y'=y(0)$ and $y''=y(T)$, for
the $\upsilon$-adic classical action we obtain
\begin{equation}
  \label{classact}
\bar
S_\upsilon(y'',T;y',0)=\frac{\sqrt{mg\beta}}{2\sinh(\sqrt{\frac{g\beta}{m}
} T)}\left[(e^{\frac{2\beta}{m}y'}+e^{\frac{2\beta}{m}y''})\cosh
(\sqrt{\frac{g\beta}{m}} T)-2e^{\frac{\beta}{m}(y'+y'')}\right].
\end{equation}

In the $p$-adic case, we get a constraint which arises from the
investigation of the domain of a convergence analytical function
which appears during the derivation of the formulae (\ref{gensol}).
This constraint is $|\dot y|_p\leq \frac{1}{p}|\sqrt{\frac{g
m}{\beta}}|_p$.

By the transformation $X=\frac{m}{\beta}e^{\frac{\beta} {m}y}$, we
can convert Lagrangian (\ref{eq:eq3}) in a more suitable, quadratic
form
\begin{equation}
  \label{eq:eq6}
  L(X,\dot X)=\frac{m\dot X^2}{2} +\frac{g\beta X^2}{2}.
\end{equation}

For the conditions $X'=X(0)$, and $X''=X(T)$, action for the
classical $\upsilon$-adic solution $\bar X(t)$ is
\begin{equation}
  \label{lj3}
\bar S_\upsilon(X'',T;X',0)= \frac { \sqrt{mg\beta} } { 2\sinh(\sqrt
{\frac{g\beta}{ m}}T)} \left[(X'^2+X''^2) \cosh(\sqrt{\frac{g\beta}{
m}} T)-2X'X''\right].
\end{equation}
We note that this action is different from the action (\ref{5.4})
only in one constant term. Because action (\ref{lj3}) is quadratic
one (with respect to the initial and final point), the corresponding
kernel is \cite{goranbranko}
\begin{equation}
  \label{lj4}
  {\cal K}_\upsilon
(X'',T;X',0)= \lambda_\upsilon \left(\frac{1}{2h} \frac{\sqrt{g\beta
m}} {\sinh(\sqrt{\frac{g\beta}{ m}}
  T)}\right)
  \left|\frac{1}{h}
  \frac{\sqrt{g\beta m}}{\sinh(\sqrt{ \frac{g\beta}{ m}}
  T)}
  \right|_\upsilon^{1/2}
  \chi_\upsilon\left(-\frac{1}{h}\bar
S_\upsilon\right),
\end{equation}
where $\chi_\upsilon$ is the adelic additive character \cite{vvz}.

 The necessary condition for the existence of an adelic model
is the existence of a $p$-adic quantum-mechanical ground state
$\Omega(|X|_p)$, i.e.
\begin{equation}
\label{6.1} \int_{|X'|_p\leq1}{\cal K}_p (X'',T;X',0)dX'=
\Omega(|X''|_p).
\end{equation}
Analogously, if a system  is in the state $\Omega(p^\nu|X|_p)$, then
its kernel must satisfy
\begin{equation}
\label{6.2} \int_{|X'|_p\leq p^{-\nu}}{\cal K}_p
({X}'',N;{X}',0)d{X}'= \Omega(p^\nu|X''|_p).
\end{equation}
In case the the $p$-adic  ground state is of the form of the
$\delta$-function, we have to investigate conditions under which
the corresponding kernel of the model  satisfies the equation
\begin{equation}
\label{6.3} \int_{Q_p}{\cal K}_p(X'',T;X',0)\delta(p^\nu-|X'|_p)dX'=
\chi_p(ET)\delta(p^\nu-|X''|_p),
\end{equation}
with zero energy $E=0$. In what follows, we apply (\ref{6.1}),
(\ref{6.2}) and (\ref{6.3}) to our model. As a result for the
$p$-adic wave functions (in the case $p\neq 2$), we get
\begin{equation}
\label{wf1} \Psi_p(X)=\Omega(|X|_p), \quad |T|_p\leq
\left|\frac{m}{2h}\right|_p,\quad \left|\frac{g\beta
m}{4h^2}\right|_p<1
\end{equation}
\begin{equation}
\label{wf2} \Psi_p(X)=\Omega(p^\nu|X|_p), \quad |T|_p\leq
\left|\frac{m}{2h}\right|_p p^{-2\nu},\quad \left|\frac{g\beta
m}{4h^2}\right|_p\leq p^{3\nu}
\end{equation}
\begin{equation}
\label{wf3} \Psi_p(X)=\delta(p^\nu-|X|_p), \quad
\left|\frac{T}{2}\right|_p\leq \left|\frac{m}{h}\right|_p
p^{2\nu-2},\quad \left|\frac{g\beta m}{h^2}\right|_p\leq
p^{2-3\nu}.
\end{equation}

The above conditions are in accordance with the conditions for the
convergence of the $p$-adic analytical functions which appear in the
solution of the equation of motion (\ref{gensol}) and the classical
action (\ref{classact}). We see there is a wide freedom in choosing
the parameters of the model, such as mass of the tachyon field $m$,
damping factor $\beta$, parameter $g$ related to the ``strength of
the constant gravity``, and cosmological constant $\Lambda$ which
appears in the de Sitter $(2+1)$ dimensional model. A relevant
physical conclusion served from these relations still needs a more
realistic model with tachyon matter and with a precise form of
metrics.

\section{Conclusion}

In spite of the very attractive features of the tachyonic
inflation, first of all, the rolling tachyon condensate, this
approach faces difficulties such as reheating \cite{sami}. It
seems that both mechanisms, based on real tachyons - conventional
reheating mechanism and quantum mechanical particle production
during inflation - do not work. Recent results in nonlocal
($p$-adic) tachyon inflation \cite{barnaby,jokuovskaya}, in which
a $p$-adic tachyon drives a sufficiently long period of inflation
while rolling away from the maximum of its potential deserve much
more attention. The {\it classical} $p$-adic models succeed with
inflation where the real string theory fails. In this paper we
have calculated a {\it quantum} propagator for the $p$-adic and
adelic tachyons, found conditions for the existence of the vacuum
state of $p$-adic and adelic tachyons, noted interesting relations
with the minisuperspace closed homogenous isotropic model in
$(2+1)$ dimensions using Einstein gravity with a cosmological
constant and an antisymmetric tensor field matter source
\cite{we-vol,we-myers}. We have shown that the new results can
give rise to a better understanding of the $p$-adic and real
quantum tachyons, their relation via Freund-Witten formula and a
possible role of tachyon field as a dark matter. Our results can
also be used as a basis for further investigation of ($p$-adic)
quantum mechanical damped systems and corresponding wave functions
of the universe in the minisuperspace models based on the
tachyonic matter with different potentials. Further investigation
should contribute to the better understanding of quantum rolling
tachyon scenario in a real \cite{AmbjornJanik} and $p$-adic case.

\section*{Acknowledgement}
This work is partially supported by the Ministry of Science of the
Republic of Serbia under Grants 144014 and 141016. Work on this
paper is also supported in part by the UNESCO-BRESCE/IBSP grant No.
875.854.7, within the framework of the Southeastern European Network
in Mathematical and Theoretical Physics (SEENET-MTP).

\end{document}